\documentstyle[fl eqn,twoside]{article}
\def\be{\begin{equation}}
\def\ee{\end{equation}}
\def\bi{\bibitem}
\begin{document}
\title{Study of symmetry in F(R) theory of gravity}
\author{Abhik Kumar Sanyal}
\maketitle \noindent
\begin{center}
Dept. of Physics, Jangipur College, Murshidabad, \noindent
India - 742213\\
\noindent
e-mail : sanyal$_{-}$ak@yahoo.com\\
\end{center}
\noindent
\date{}
\maketitle
\begin{abstract}
An action in which the Ricci scalar is nonminimally coupled with a 
scalar field and contains higher order curvature invariant terms
carries a conserved current under certain conditions that decouples 
geometric part from the scalar field. The conserved current relates 
the pair of arbitrary coupling parameters $f(\phi)$ and $\omega(\phi)$ 
with the gravitational field variable, where $\omega(\phi)$ is the
Brans-Dicke coupling parameter. The existence of such conserved
current may be helpful to sketch the cosmological evolution from
its early age till date in a single frame.
\end{abstract}
\section{\bf{Introduction}}
In the recent years, modified theory of gravity is being treated
all around with increasing importance, with the anticipation that
such a theory of gravity which includes higher order curvature
invariant terms  may help in the understanding of the dark energy
origin (for a review, see \cite{n:r}). In this connection, various
modified $F(R)$ theories of gravity have been constructed and applied to the
description of the late-time cosmic acceleration together with the check for
local tests (see eg., \cite{n:o} and the references therein) . 
$F(R)$ theory originated in the early sixties with
the work of Utiyama and B.De-Witt \cite {r:b}. They demonstrated
for the first time that for renormalizability at one-loop level,
the classical action should be supplemented by higher order
curvature invariant terms. Later, the search for a viable
semiclassical or quantum model for the dark energy component and
its dynamical properties has turn out to be one of the most
important topics of string cosmology, which attempts to reveal the
cosmological implications of string effective actions. In
particular, quantum corrections to the Einstein-Hilbert action
naturally arise for the closed string, and modify the
gravitational interaction in a way which might be testable in the
near future. More recently, there has been many calculations
showing that, when quantum corrections or String/M-theory are
considered, the effective classical gravitational action admits
higher order corrections in curvature invariants
\cite{l:s} - \cite{d:v}.\\
Higher-order gravity terms can generate non-trivial effects in a
four-dimensional spacetime. If the dilaton varies in spacetime,
gradient terms appear and higher order curvature invariant terms
interact with a non-constant $\alpha'$-coupling. Also,
compactification of the $26$ or $10$ - dimensional target space is
encoded in residual modulus fields which, in general, will evolve
in time. Both the cases finally yield a dynamical field
non-minimally coupled to gravity. Dilaton and modulus cosmologies
in higher order gravity were considered, for example, in
\cite{m:s} - \cite{p:y}. In connection with the recent
observations, dark energy models with higher order curvature
invariant terms were inspected, in the case of fixed moduli
\cite{s:t}, while a dynamical dilaton
as a dark energy candidate was studied in \cite{g:p, p:t}. \\
Although $F(R)$ theories offer a chance to explain the
acceleration of the universe, they are not free of problems. For
instance, the application of their metric formulation to the
homogeneous and isotropic Robertson-Walker geometry yields a
fourth order nonlinear coupled (with the scalar field)
differential equation for the scale factor $a(t)$, which in general
cannot be analytically solved even for the simplest form of
$F(R)$. Symmetry plays an important role to find analytical
solutions, eg., it has been possible to find exact analytical
solutions for $R^2$ gravity, by invoking N\"other symmetry
\cite{a:k}. Further, it has been shown \cite{a:s}, \cite{a:ks},
that there exists a conserved current, other than N\"other
current, for a general scalar tensor theory of gravity,
nonminimally coupled to a scalar field under certain conditions.
The use of such conserved current in tackling the field equations
has also been demonstrated in \cite{a:ks}. Additionally, it has
also been shown \cite{a:ks} that the same conserved current exists
even for higher order theory of gravity, under a different
condition. The importance of such symmetry had not been expatiated
there. In the present work, we therefore would like to demonstrate
the use of such symmetry in the context of exploring exact
analytical solutions of higher order theory of gravity. It has
been shown that under the assumption on the existence of such
symmetry, the fourth order differential equation decouples scalar
field from the gravitational field variables. Thus, it becomes
less difficult to tackle the field equations. The study with a maximally
symmetric formalism in $F(R)$ theory of gravity in the background of
homogeneous and isotropic space time has been taken up recently \cite{tm}.
However, the symmetry under consideration in the present work is
independent of background space-time and the action has got a more
general form.\\
In the following section we review our
earlier work \cite{a:ks}, ie., find the conserved current for a
general action corresponding to $F(R)$ theory of gravity. The action
under consideration does not incorporate Gauss-Bonnet term
in $4$-dimensional space-time. Nevertheless, adding terms of the type
$R^n$ is no less important, since they can produce early times
inflation \cite{aa}, and late time cosmic acceleration \cite{sv}, 
\cite{nod}.
In section 3, we take up $R^2$ theory of gravity as an example to
demonstrate how it works suitably to explain cosmological
evolution from early Universe till date.

\section{\bf{Conserved current for an action containing higher order
curvature invariant terms}}

The $4$-dimensional string effective action under loop level
approximation \cite{j:h, c:j} containing additional higher order
curvature invariant terms can be expressed as,

\be
A = \int\left[f(\phi) ~R + B F(R)
-\frac{\omega(\phi)}{\phi}\phi,_{\mu}\phi^{,\mu} -V(\phi)-\kappa
L_{m}\right]\sqrt{-g}~d^4x, \ee where, $F(R)$ is an arbitrary
function of Ricci scalar, $B$ is the coupling constant and $L_{m}$
is the matter field Lagrangian, while $f(\phi)$ and $\omega(\phi)$
are coupling parameters, the later is of Brans-Dicke \cite{bd} origin.
Field equations corresponding to the above action are,

\[
f\left(R_{\mu\nu}-\frac{1}{2}g_{\mu\nu}R\right)+f^{;\alpha}_{;\alpha}g_{\mu\nu}-f_{;\mu
;\nu} -\frac{\omega}{\phi}\phi_{,\mu}\phi_{,\nu}+
\frac{1}{2}g_{\mu\nu}\left(\frac{\omega}{\phi}\phi_{,\alpha}\phi^{,\alpha}+V(\phi)\right)\]
\be +B\left[(F,_R)R_{\mu\nu}-\frac{1}{2}F g_{\mu\nu}+
{{(F,_R)}^{;\alpha}}_{;\alpha}~ g_{\mu\nu}-(F,_R)_{;\mu
;\nu}\right]=\frac{\kappa}{2}T_{\mu\nu} \ee

\be
Rf'+2\frac{\omega}{\phi}\phi^{;\mu}_{;\mu}+\left(\frac{\omega'}{\phi}
-\frac{\omega}{\phi^2}\right)\phi^{,\mu}\phi_{,\mu}-V'(\phi) = 0
\ee In the above, $F,_R$ denotes derivative of the function $F(R)$
with respect to $R$ and $T_{\mu\nu}$ is the energy momentum tensor
corresponding to the matter field (barotropic fluid) Lagrangian, $L_{m}$.
It is practically impossible to solve the above set of field equations
even in the homogeneous and isotropic background geometry. In the
homogeneous and isotropic background geometry, these are only a pair of
independent field equations, with at least five field variables
(if $F(R)$ is known a-priori), viz., $a, \phi, f(\phi), \omega(\phi)$
and $V(\phi)$, in the early Universe, and more $(\rho, p)$, in the
radiation and matter dominated era. Thus, not only it is difficult
to solve the above set of field equations due to the presence of
fourth order coupled nonlinear differential equations, but also it
requires at least three additional equations (assumptions) to obtain
exact analytical solutions. One of the assumptions may be invoking the
presence of dynamical symmetry, since it turns out to be a very powerful
tool in this regard. It decouples the equations at one hand and gives
additional equations, to make things a little bit tractable. In view of
this discussion, let us now proceed to find the condition required to
have some dynamical symmetry of the theory, directly from the field equations.\\
The trace of equation (2) is the following,

\be
Rf-3f^{;\mu}_{;\mu}-\frac{\omega}{\phi}\phi^{,\mu}\phi_{,\mu}-2V
-B[R(F,_R)+3{{(F,_R)}^{;\alpha}}_{;\alpha}-2F] =
\frac{\kappa}{2}T^{\mu}_{\mu}. \ee Now since,

\[f^{;\mu}_{;\mu} = f''\phi^{,\mu}\phi_{,\mu} + f'\phi^{;\mu}_{;\mu},\]
so, multiplying equation (3) by $f$ and equation (4) by $f'$ and eliminating
$Rff'$ between the two, we obtain,

\be
\left[(3f'^2+2f\frac{\omega}{\phi})^{\frac{1}{2}}\phi^{;\mu}\right]_{;\mu}+
\left(3f'^2+2f\frac{\omega}{\phi}\right)^{-\frac{1}{2}}\left[B
f'[R(F,_R) +3{{(F,_R)}^{;\alpha}}_{;\alpha}-2F]
+\frac{\kappa}{2}f'~T^{\mu}_{\mu}
-f^3\left(\frac{V}{f^2}\right)'\right]=0. \ee In view of equation
(5) we can conclude that under the following condition,

\be B [R(F,_R) +3{{(F,_R)}^{;\alpha}}_{;\alpha}-2F] = -
\frac{\kappa}{2}~T^{\mu}_{\mu} +
  \frac{f^3}{f'}\left(\frac{V}{f^2}\right)', \ee
there exists a conserved current $J^\mu$,

\be
J^{\mu}_{;\mu}=[(3f'^2+2f\frac{\omega}{\phi})^{\frac{1}{2}}\phi^{;\mu}]_{;\mu}
= 0, \ee corresponding to the general form of $F(R)$ theory of
gravitational action (1). In the case of homogeneous cosmology,
left hand side of condition (6) contains fourth order time
derivative of the scale factor $a$, as it should be, while the
right hand side is an ordinary function of $\phi$, during the
early Universe and the radiation dominated era, since, $T^\mu_\mu$
vanishes at those epoch. In the matter dominated era ($p = 0$), an
additional functional dependence of the scale factor $a$ appears on
the right (since $\rho = \frac{\rho_0}{a^3}$, $\rho_0$ being a constant).
Further the expression for the conserved current $J^\mu$ given in (7)
takes the following simplified form in the case of homogeneous cosmology,

\be
\sqrt{\left(3f'^2+2f\frac{\omega}{\phi}\right)}\;\;\dot \phi
= \frac{c}{\sqrt {-g}}, \ee where, $c$ is a non-vanishing constant 
and $g$ is the determinant of the metric. Hence, the
cosmological equations corresponding to $F(R)$ theory of gravity
becomes somewhat easier to tackle. In addition, the last term on
the right hand side of condition (6) vanishes either for dilatonic
scalar, $V(\phi) = 0$, or if the potential is proportional to the
square of the coupling parameter $f(\phi)$. Therefore, in general,
one can conclude that the following action corresponding to $F(R)$
theory of gravity,

\be
A = \int\left[f(\phi) ~R + B F(R)
-\frac{\omega(\phi)}{\phi}\phi,_{\mu}\phi^{,\mu} -V(\phi) -\kappa L_{m}\right]\sqrt{-g}~d^4x, \ee
admits the integral of motion (7), under the condition

\be B [R F,_R +3{{(F,_R)}^{;\alpha}}_{;\alpha}-2F] = -
\frac{\kappa}{2}~T^{\mu}_{\mu},  \ee
provided,

\be V = \lambda f^2,\ee
where, $\lambda$ is a constant, which
vanishes for dilatonic scalar. It is interesting to note that
the left hand side of condition (10), under which the conserved
current (7) exists, contains only geometrical part and is overall
free from the scalar field. It is also to be noted that so far we have
made two assumptions, the existence of symmetry and a relation
between $f(\phi)$ and $V(\phi)$. The first assumption yielded a
couple of additional equations viz., (7) and (10). So, altogether
we now have four equations between (2), (3), (7) and (10), out of
which only three are independent, since (7) and (10) are derived
in view of equations (2) and (3). So at the end, we observe that
the existence of symmetry, decouples the scalar field from the
geometry, increasing the possibility of obtaining analytical
solutions of the field equations. Additionally, instead of the
very complicated form of equation (2), one can now use three
considerably simpler equations out of the four, viz., (3), (4),
(7) and (10). In fact equation (4) takes further simplified form
in view of equation (10), as

\be
Rf-3f^{;\mu}_{;\mu}-\frac{\omega}{\phi}\phi^{,\mu}\phi_{,\mu}-2V = 0,\ee
where, $V = \lambda f^2$, as already mentioned. Thus, under different
choices of the function $F(R)$, analytical solution may be found just
by solving the fourth order nonlinear differential equation (10) of the
scale factor $a$, in the case of Robertson-Walker metric. One more
assumption is required to find the form of $f(\phi), \omega(\phi)$,
and $\phi = \phi(t)$, in view of other pair of independent equations
(8), and (12). In the following section we cite a nontrivial example
to demonstrate the use of such symmetry.

\section{Cosmic evolution with $R^2$ term}
The exact solutions of the fourth order gravity equations
may sometimes be found considerably easily if we consider the
existence of a conserved current given in equation (7). To establish the
importance of the conserved current, let us consider homogeneous and
isotropic cosmological model described by the Robertson-Walker metric,

\be ds^2 = - dt^2 + a^2(t) [\frac{dr^2}{1-kr^2} + r^2 (d\theta^2 +
sin^2 \theta d\phi^2)],\ee
for which the Ricci scalar is given by,

\be R = 6\left(\frac{\ddot a}{a} + \frac{\dot a^2}{a^2} +
\frac{k}{a^2}\right). \ee
If we now take as an example, $F(R) = R^2$, Then, equation (10),
which is the condition for the existence of conserved current
reads,

\be R^{;\mu}_{;\mu} = \Box R = -(\ddot R + 3\frac{\dot a}{a} R)  =\frac{\kappa}{2} T^{\mu}_{\mu}. \ee

\noindent
\textbf{Early Universe :}\\

\noindent
In the early Universe, $\rho = p = 0$, ie., $T^{\mu}_{\mu} = \rho - 3p = 0$,
and so, $ \Box R = 0$, in view of equation (15), ie.,

\be \ddot R + 3\frac{\dot a}{a} R  = 0. \ee The above equation
admits exponential solution for $k = 0$, and a solution in the
form, $a = a_{0} t$, for $k \ne 0$. In the second case, the
Universe undergoes just the amount of inflation required to solve
the problems of isotropy and homogeneity, irrespective of the form
of the potential, which may be zero as in the case of dilatonic
scalar. Further, there is also no need of slow roll approximation.
One has to make yet another choice at  this stage to find other
field variables, viz, $f(\phi), \omega(\phi)$ and $\phi(t)$. For
example, one can choose a non-zero form of the potential
$V(\phi)$, to get $f(\phi)$. Thus $\omega(\phi)$ and $\phi = \phi(t)$
may be found in view of equations (8) and (12).\\

\noindent
\textbf{Radiation dominated era :}\\

\noindent
Equation (16) also holds in the Radiation dominated era as well, since,
$T^{\mu}_{\mu} = \rho - 3p = 0$. Now, if inflation is sufficient
to make the Universe spatially flat, ie., $k = 0$, then, the above
equation (16) admits the solution,

\be a = a_0 t^{\frac{1}{2}}.\ee
Thus, the Radiation era of Friedmann Universe remains unaltered even
in the higher order theory of gravity . One can now find the matter
density $\rho$, as $\rho a^4 = \rho^0$, $\rho^0$ being a constant
and so the pressure, $p = \frac{\rho}{3}$ may be evaluated. Further,
Using the same form of $f(\phi)$ as fixed in the early Universe
the form of $\omega(\phi)$, with a different evolution history
of $\phi(t)$ may be found in view of equations (8) and (12).\\

\noindent
\textbf{Matter dominated era :}\\

\noindent
In the matter dominated era, $p = 0$ and so, equation (15) reads,
$\Box R = \frac{\kappa}{2}T^{\mu}_{\mu} = \frac{\kappa}{2} \rho$.
However, since,

\be\rho a^3 = \rho_0, \ee
where, $\rho_0$ is a constant, so equation (15) finally takes the
following form,

\be \ddot R + 3 \frac{\dot a}{a}\dot R = -
\left(\frac{\kappa\rho_0}{12B}\right)\frac{1}{a^3}.\ee In the
isotropic and homogeneous case under consideration, the above
equation translates to ,

\be a^2 \stackrel\cdots a + a\dot a\ddot a -2\dot a^3 =
-\left(\frac{\kappa\rho_0}{72B}\right)t+l,\ee where $l$ is a
constant of integration. Equation (20) admits a solution in
the form,

\be a = a_0 t^{\frac{4}{3}}, \ee provided, $\rho_0  = \frac{320
B a_0 ^3 }{\kappa}$ and $l = 0$. Hence the Universe undergoes an
accelerating phase in the matter dominated era, as present
cosmological observations suggest. $\rho$ can now be found in
view of equation (18) and as discussed above a different form of
$\omega(\phi)$ and $\phi = \phi(t)$ may be found taking the same
form of $f(\phi)$ fixed in the early Universe.

\section{\bf{Concluding remarks}}
The existence of a general form of symmetry and the corresponding
conserved current for a general action corresponding to $F(R)$ theories
of gravity has been found in an earlier work \cite{a:ks}. That the
existence of the integral of motion in the cosmological context
makes it easier to handle the field equations for studying exact
solutions has been demonstrated here in the context of $R^2$ term.
In this connection the following points are noteworthy.\\
1. Action (1) excludes Gauss-Bonnet term, since it is topologically
invariant in $4$-dimensional space-time. Nevertheless, the procedure
followed in section (2) may be extended in higher dimensional
space-time to incorporate such term.\\
2. One can relax condition (11) and use equation (6) instead. In
that case, one may choose a functional $\phi$ dependent form of the
last term on the right hand side of equation (6), which may be found
for power law or exponential type of inflation in the early Universe.\\
3. In principal it may be possible to find a fixed form of $f(\phi)$
and $\omega(\phi)$, which would satisfy all the field equations from
early Universe through matter dominated era with different cosmological
evolution of $\phi(t)$.\\
4. Usually, early time inflation and late time cosmic acceleration are 
incorporated in modified theory of gravity. That the radiation dominated 
era of the Friedmann Universe remains unchanged even in such theory, is
an additional outcome of the present work.

\end{document}